\documentclass[journal]{IEEEtran}
\usepackage{array}
\usepackage{url}
\usepackage{amsmath}
\usepackage{amsthm}
\usepackage{amssymb}
\usepackage{graphicx}
\usepackage{cite}
\usepackage{subfig}
\usepackage[colorlinks = true, linkcolor=red, citecolor=red,urlcolor=blue]{hyperref}
\usepackage{multirow}
\usepackage{balance}
\usepackage{soul}

\graphicspath{{./figures/}}

\providecommand{\tabularnewline}{\\}
\theoremstyle{plain}

\theoremstyle{plain}

\providecommand{\propositionname}{Proposition}
\providecommand{\theoremname}{Theorem}

\begin{document}
	
\title{Digital Divide and Social Dilemma \\of Privacy Preservation}

\author{Hamoud~Alhazmi, Ahmed~Imran, and Mohammad~Abu~Alsheikh
\thanks{H.~Alhazmi, A.~Imran, and M.~Abu~Alsheikh are with the Faculty of Science \& Technology, University of Canberra, Australian Capital Territory, Australia. This work was supported in part by the Australian Research Council (ARC) under Grant DE200100863. The data collection in this project has been approved by the Human Research Ethics Committee at the University of Canberra under application 4522 ``Privacy Coupling: When Your Personal Devices Betray You''.}
}

\maketitle

\begin{abstract}

While digital divide studies primarily focused on access to information and communications technology (ICT) in the past, its influence on other associated dimensions such as privacy is becoming critical with a far-reaching impact on the people and society. For example, the various levels of government legislation and compliance on information privacy worldwide have created a new era of digital divide in the privacy preservation domain. In this article, the concept ``digital privacy divide (DPD)'' is introduced to describe the perceived gap in the privacy preservation of individuals based on the geopolitical location of different countries. To better understand the DPD phenomenon, we created an online questionnaire and collected answers from more than 700 respondents from four different countries (the United States, Germany, Bangladesh, and India) who come from two distinct cultural orientations as per Hofstede's individualist vs. collectivist society. However, our results revealed some interesting findings. DPD does not depend on Hofstede's cultural orientation of the countries. For example, individuals residing in Germany and Bangladesh share similar privacy concerns, while there is a significant similarity among individuals residing in the United States and India. Moreover, while most respondents acknowledge the importance of privacy legislation to protect their digital privacy, they do not mind their governments to allow domestic companies and organizations collecting personal data on individuals residing outside their countries, if there are economic, employment, and crime prevention benefits. These results suggest a social dilemma in the perceived privacy preservation, which could be dependent on many other contextual factors beyond government legislation and countries' cultural orientation.

\end{abstract}

\section*{\textbf{Introduction}}\label{sec:intro}

Over the recent years, there has been an increased interest in the digital divide, global technology change, and social inequalities. The digital divide limits opportunities for disadvantaged communities and increases social, digital, and economic inequalities. For example, the recent outbreak of the COVID-19 pandemic has exposed acute digital divide problems in telehealth access~\cite{ramsetty2020impact} and online educational technologies~\cite{lai2021revisiting}. Research and policies in the digital divide have shifted from focusing purely on the physical access of technologies (first-level digital divide), to the importance of digital literacy and skills (second-level digital divide), and recently to the positive and adverse outcomes of digital access (third-level digital divide)~\cite{scheerder2017determinants}.

The proliferation of information and communications technology (ICT) has produced a new type of digital divide  related to information privacy. At a rate and scale unforeseen just a few years ago, mobile devices and the Internet of things (IoT) track and sense data continuously on our everyday activities and life habits including our visited locations, shopping behavior, selfies, fitness, daily routines, and voice commands, just to name a few. The data includes private information and biometric records which are collected (explicitly and implicitly) by many crowdsensing services. Numerous industries rely on IoT and crowdsensing to optimize their offered services including insurance, electricity supply, public safety, logistics, telecommunications, etc. It is therefore critical to understand the digital privacy divide~(DPD) and the levels of privacy preservation provided to individuals by their governments.

To understand the emerging problem of DPD, this article addresses the problem of DPD and analyzes the gap in privacy preservation of individuals residing at different parts of the world (the United States, Germany, Bangladesh, and India). We collected more than 700~responses on how individuals perceive the levels of privacy protection provided by their governments when accessing ICT systems. We first operationalize the DPD concept based on the well-established privacy rights of the General Data Protection Regulation (GDPR)~\cite{gdpr2016general} defined by the European Parliament and Council of the European Union. Accordingly, we designed a questionnaire consisting of 11 questions divided into 3 categories. Category~A includes 8~questions (Q1-Q8) to define the privacy levels provided by the governments of the four countries. Category~B consists of 2 questions (Q9 and Q10) that address the social dilemma in privacy, e.g.,~how people perceive the importance of their privacy and the privacy protection of others. Finally, Category~C has one question (Q11) to collect open-ended feedback from the respondents. We also used 3 attention checks~(A1-A3) and 3 timing checks~(T1-T3) for quality assurance of the response collection and to ensure reliability and validity of collected data.

We have two major findings. First, we find that the privacy concerns of individuals cannot be justified using theories of popular cultural dimensions, e.g.,~the Hofstede's model~\cite{hofstede2011dimensionalizing}. Respondents residing in Germany and Bangladesh have expressed more concerns overall on their privacy compared to those residing in the United States and India (Q1-Q8). Second, we find that most of the respondents (80\%, 81\%,  74\%, and 80\% of the respondents in the United States, India, Bangladesh, and Germany, respectively) think governments should enforce restrict privacy policies on domestic companies and organizations that collect their personal data in all cases, regardless of the economic, employment, and crime prevention benefits of the data collection (Q10). On the contrary, 66\%, 69\%, 70\%, and 44\% of the respondents in the United States, India, Bangladesh, and Germany think that governments should allow domestic companies and organizations collecting personal data on individuals residing outside their countries if there are economic, employment, and crime prevention benefits (Q9).

The rest of this article is organized as follows. We first present an overview of information privacy in ICT systems and its relation to the DPD problem. Then, we discuss the operationalization of the DPD concept into abstract privacy concepts and the design of the survey research (questionnaire). After that, we discuss the findings of the DPD questionnaire and the social dilemma of privacy. We also suggest recommendations for closing the DPD gap. Finally, we outline some interesting research directions and  conclude the article.
\section*{\textbf{Digital Privacy Divide (DPD)}}\label{sec:sec_1}

\begin{figure*}
	\begin{centering}
		\includegraphics[width=0.98\textwidth,trim=1cm 1cm 1cm 1cm]{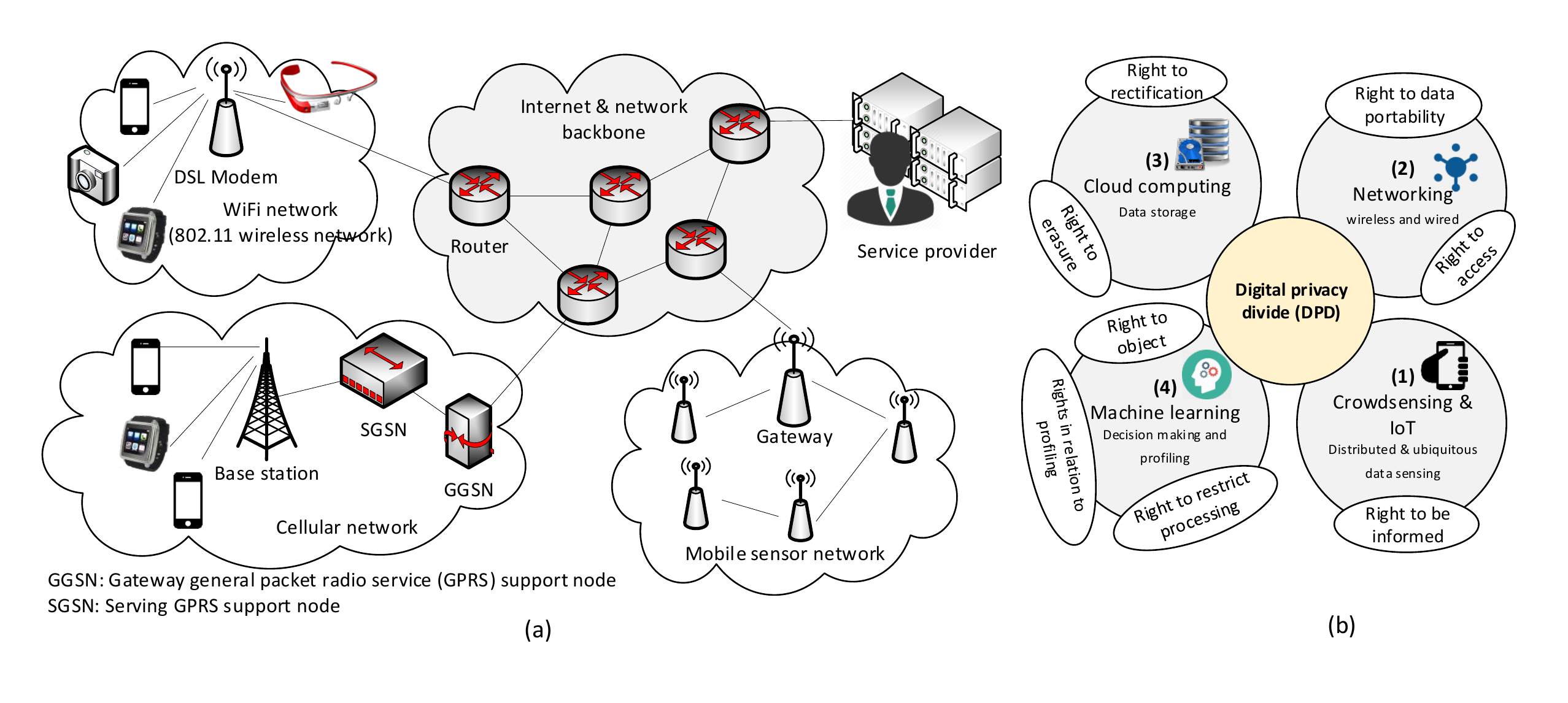}
		\par\end{centering}
	
	\caption{Illustration of possible data access points and associated DPD issues. (a)~Modern ICT networks. (b)~Key ICT advances behind the DPD era.\label{fig:system_model}}
\end{figure*}

This section first introduces the pervasive data collection and processing in ICT systems and discusses its corresponding privacy concerns.

\subsection*{\textbf{Pervasive ICT services}}

The proliferation of ICT services is enabled by several emerging technologies in IoT for data sensing, high-speed networks for data transmission, cloud computing for data storage and processing, and machine learning for data analytics. Figure~\ref{fig:system_model}~(a) shows an example of a modern ICT network. IoT devices are embedded with various sensors, e.g.,~GPS, cameras, accelerometers, and microphones, that sense data about people and their surroundings. The data is transmitted through high-speed wireless, e.g.,~5G, and wired, e.g., fiber optics, networks to service providers. Service providers use efficient cloud computing, e.g.,~Amazon Web Services, to store and process the data. Machine learning is applied to learn the relationships in data and create new services, e.g.,~real-time traffic monitoring, speech recognition, and activity recognition.

\subsection*{\textbf{Privacy rights}}

Next, we discuss the privacy rights defined in the general data protection regulation (GDPR)~\cite{gdpr2016general}. Unlike other privacy regulations, we find that the GDPR provides well-defined and up-to-date privacy rights. There are eight GDPR privacy rights that provide individuals with control over their data.

\begin{itemize}
	\item \textit{Right to be informed}: The right to know who is collecting and accessing their data.
	\item \textit{Right of access}: The right to access their personal data.
	\item \textit{Right to rectification}: The right to request correcting inaccurate records of their personal data.
	\item \textit{Right to erasure}: The right to be forgotten by deleting their data and preventing future data collection.
	\item \textit{Right to restrict processing}: The right to restrict the processing of specific categories of their personal data.
	\item \textit{Right to data portability}: The right to transfer their data to specific recipients of their choice.
	\item \textit{Right to object}: Individuals have the right to provide and withdraw consent on the processing and collection of their personal data.
	\item \textit{Rights in relation to automated decision-making and profiling}: The right to opt-out of the use of their data in automated systems including machine learning and artificial intelligence~(AI).	
\end{itemize}

\subsection*{\textbf{The era of DPD}}


We are witnessing the emergence of the DPD era due to the various levels of protection and legislation available to individuals while accessing ICT services. Figure~\ref{fig:system_model}~(b) shows the main ICT components related to the DPD problem and the corresponding privacy rights that must be met at each component.

\subsubsection*{\textbf{DPD in crowdsensing and IoT}}

Crowdsensing and IoT devices, such as wearables, mobile phones, wrist bands, and fitness trackers, collect massive amounts of data on people including vital signs, movements, visited locations, shopping habits, and daily activities. 

When using crowdsensing and IoT for data collection, the users must be provided with sufficient information on the data sensing activities, the length of data retention, the aims of data sensing, and who will access the data. Service providers must provide individuals with clear privacy details before collecting any personal data. For example, the authors in~\cite{kement2021holistic} showed that privacy breaches, e.g.,~appliance usage inference, can occur when using IoT for smart electricity, water, and natural gas metering. Providing incomplete information on possible privacy breaches to IoT and crowdsensing users will widen the DPD gap.

\subsubsection*{\textbf{DPD in networking}}
High-speed networks, e.g.,~5G and fiber optics, are vital for the operations of ICT services. 

Individuals must be able to gain access to their personal data that has been collected by service providers. Therefore, service providers must be connected to high-speed networks, such that individuals can download copies of their personal data (which can be multiple gigabytes). Moreover, users must have the right to data portability. Users must have the ability to transfer their personal data to third-party recipients of their choice. To meet these requirements, service providers must use high-speed networking technologies for data portability and access.

The failure to provide high-speed networking for both data portability and access will contribute to increasing the DPD gap.

\subsubsection*{\textbf{DPD in cloud computing}}

Service providers must implement cloud application programming interfaces (cloud APIs) that enable the right to rectification. They allow users to correct their personal data if it is inaccurate. Furthermore, if the personal data is incomplete, users should be allowed to request completing the data collection. 

The right to erasure known as (right to be forgotten) must be also enabled in cloud APIs. It allows users to request service providers to delete their personal data. Organizations should erase the data if (a)~the personal data is not needed for the original purpose, (b)~a user withdraws consent, or (c)~the data is processed unlawfully.

The failure to provide accessible cloud APIs for data rectification and erasure will widen the DPD gap. 

\subsubsection*{\textbf{DPD in machine learning}}

Machine learning, e.g.,~deep learning, is widely adopted for extracting hidden patterns from the collected data. New types of privacy attacks have recently emerged by exploiting the modeling parameters stored in the machine learning models~\cite{liu2021machine}. This means that even though the adversary may not directly access the actual personal data, they can still compromise the privacy of individuals by exploiting the prediction output of ICT services. For example, membership attacks can be exploited to reveal the health conditions of individuals in e-health services.

Users must have the right to object to the processing of their personal data, depending on the objectives and the lawful basis of processing. Moreover, users must be able to restrict the processing and limit the use of their personal data. For example, users may decide to restrict the data processing when keeping the data for legal purposes or while the service provider is verifying the request for data erasure. Finally, users have privacy rights in relation to automated decision-making with machine learning, e.g.,~profiling users based on their health, work performance, and personal interests without human intervention. When applied in ICT services, machine learning programs must be implemented to support the aforementioned privacy rights. Machine learning programs, which do not comply with the processing requirements, contribute to increasing the DPD gap.

\section*{\textbf{DPD Questionnaire}}\label{sec:sec_2}

Developing an online questionnaire to identify and measure the DPD gap is not simple. Self-reported responses of questionnaires have limitations, including biased over-reporting, under-reporting, misunderstanding questions, and social desirability, i.e.,~providing responses that are favorable by the social group. These challenges are tackled by following well-established, rigorous question development processes, e.g.,~operationalization, validation, pilot studies, and quality checks~\cite{redmiles2017digital,saris2014design}. This section discusses the process of operationalizing the DPD concept into legitimate survey questions for the target audience. We first define the operationalization links between the DPD concept and the basic GDPR privacy rights. Then, we will discuss the basic assertions used to represent the DPD concept in our questionnaire. Finally, the choices of quality checks (attention questions) and population samples will be discussed for reliability and validity.   

\subsection*{\textbf{Choice of operationalization}}

\begin{figure*}
	\begin{centering}
		\includegraphics[width=0.98\textwidth,trim=1cm 0cm 1cm 1cm]{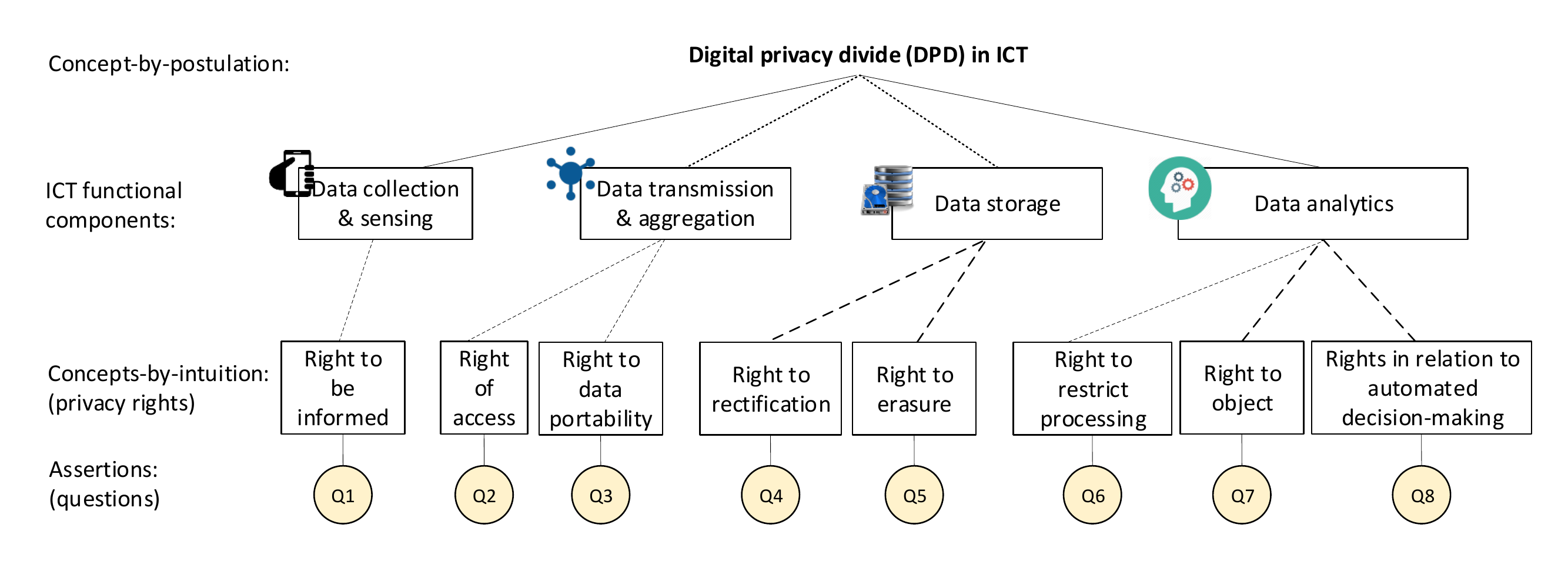}
		\par\end{centering}
	
	\caption{Operationalization of the DPD concept in ICT. Q1-Q8 are the questions of the DPD questionnaire.\label{fig:operationalization}}
\end{figure*}

DPD is a complex concept which is difficult to measure using a simple and direct question. We apply the operationalization method of defining the concepts-by-postulation (complex variables) of a study using concepts-by-intuition (abstract, measurable variables)~\cite{saris2014design}. Figure~\ref{fig:operationalization} shows the operationalization of the DPD concept into corresponding privacy concepts and the developed questionnaire questions that represent the key concepts of the study. The privacy rights are abstract concepts, and they can be clearly measured with questions. The questionnaire captures all functional components of a modern ICT system, including data collection and sensing (IoT and crowdsensing), data transmission and aggregation (networking), data storage (cloud computing), and data analytics (machine learning). The corresponding questions that measure all privacy rights are as follows. 

\begin{itemize}
	\item \textit{Q1}: I receive clear information on how my personal data is collected by my government, including who is accessing and processing the data, and the purposes of the data collection.
	\item \textit{Q2}: I have the ability to access copies of my personal data which has been collected by my government.
	\item \textit{Q3}: I have the ability to transfer my personal data which has been collected by my government to third-party recipients, e.g., organizations, of my choice.
	\item \textit{Q4}: I am able to correct my personal data which has been collected by my government when it contains inaccurate, invalid, or misleading data.
	\item \textit{Q5}: I have the ability to request deleting specific records of my personal data which has been collected by my government when the data is no longer needed for the original purpose.
	\item \textit{Q6}: I have the option and control to restrict the processing of specific categories of my personal data which has already been collected by my government.
	\item \textit{Q7}: I have the control and ability to grant or withdraw consents on collecting and processing my personal data by my government at any time.
	\item \textit{Q8}: I have the option and control to opt-out from using my personal data which has been collected by my government in making decisions and profiling, based solely on automated processing.
\end{itemize}

Two questions (Q9 and Q10) are also included to measure the social dilemma in privacy preservation.

\begin{itemize}
	\item \textit{Q9}: My government should allow domestic companies and organizations collecting personal data on individuals residing outside my country if there are economic, employment, and crime prevention benefits for my country.
	\item \textit{Q10}: My government should enforce restrict privacy policies on domestic companies and organizations that collect my personal data in all cases, regardless of the economic, employment, and crime prevention benefits of the data collection.
\end{itemize}

\subsection*{\textbf{Quality of the questionnaire}}

\begin{table}
	\caption{Quality checks applied in the web questionnaire. A1-A3 are attention checks. Correct answers are in bold. T1-T3 are time checks.}\label{tab:quality_checks}
	\begin{tabular}{|>{\centering}p{0.15\columnwidth}|>{\centering}p{0.45\columnwidth}|>{\raggedright}m{0.25\columnwidth}|}
		\hline 
		\textbf{Type} & \raggedright{}\textbf{Quality check} & \raggedright{}\textbf{Answers}\tabularnewline
		\hline 
		\hline 
		\multirow{2}{0.15\columnwidth}{\raggedright{}Attention check with feedback} & \raggedright{}\textbf{A1}) Based on the text below, what would you say your
		favorite food is? This is a simple question. When asked for your favorite
		food, you need to select ``fruits''. & \raggedright{}$\ensuremath{\triangleright}$Vegetables $\ensuremath{\triangleright}$Meat
		$\ensuremath{\triangleright}$Cereals $\ensuremath{\triangleright}$\textbf{Fruits}
		$\ensuremath{\triangleright}$Rice$\ensuremath{\triangleright}$Desserts\tabularnewline
		\cline{2-3} \cline{3-3} 
		& \raggedright{}\textbf{A2}) It is correct to say that 2 plus 3 equals 17.  & \raggedright{}$\ensuremath{\triangleright}$Strongly agree $\ensuremath{\triangleright}$Somewhat
		agree $\ensuremath{\triangleright}$Neither agree nor disagree/ I
		do not know $\ensuremath{\triangleright}$Somewhat disagree $\ensuremath{\triangleright}$\textbf{Strongly
			disagree}\tabularnewline
		\hline 
		\multirow{1}{0.15\columnwidth}{\raggedright{}Attention check without feedback} & \raggedright{}\textbf{A3}) It is correct to say that 1 plus 2 equals 3. & \raggedright{}$\ensuremath{\triangleright}$\textbf{Strongly agree}
		$\ensuremath{\triangleright}$Somewhat agree $\ensuremath{\triangleright}$Neither
		agree nor disagree/ I do not know $\ensuremath{\triangleright}$Somewhat
		disagree $\ensuremath{\triangleright}$Strongly disagree\tabularnewline
		\hline 
		\multirow{4}{0.15\columnwidth}{\raggedright{}Time check} & \raggedright{}\textbf{T1}) Time required for answering Q1-Q8 & \raggedright{}computed until the last click\tabularnewline
		\cline{2-3} \cline{3-3} 
		& \raggedright{}\textbf{T2}) Time required for answering Q9 & \raggedright{}computed until the last click.\tabularnewline
		\cline{2-3} \cline{3-3} 
		& \raggedright{}\textbf{T3}) Time required for answering Q10 & \raggedright{}computed until the last click.\tabularnewline
		\cline{2-3} \cline{3-3}
		\hline 
	\end{tabular}
	
\end{table}

Considering the reliability and validity of survey data is of utmost importance to provide credible results. We applied three-step quality checks and filtering on the collected responses. Two of these were attention checks and one was a timing check. Table~\ref{tab:quality_checks} lists the details of the quality checks applied in the study. 

\begin{itemize}
	\item \textit{Attention checks}: A1 and A2 are attention checks with feedback, i.e.,~they warn the respondent of incorrect responses and describe the importance of accurate responses. A2 is an attention check without feedback, i.e.,~it stores the answers provided by the respondents without providing explicit feedback on wrong answers. 14.28\% of the respondents failed A3. This fail percentage is slightly higher than some previous studies, e.g.,~10\% of the respondents failed similar simple attention checks in~\cite{bonnefon2016social}.
	\item \textit{Time checks}: Several studies have discussed the relationship between speeding and response quality, e.g.,~straight-lining, in web questionnaires~\cite{zhang2014speeding}. We implemented three timers to compute the time need to answer Q1-Q8, Q9, and Q10, respectively. We excluded all responses that fall within the first percentile of timing.
\end{itemize}

To eliminate the ambiguity of survey questions, a pilot test was conducted among more than ten local respondents. Feedback from those respondents was taken  to improve the wordings, removing ambiguity, and changing the order and structure of the questions.

\subsection*{\textbf{Choice of population}}

We recruited more than 700~respondents using Amazon Mechanical Turk (MTurk) and manual sharing. We restricted the participation to respondents in the United States, Germany, Bangladesh, and India. We could not collect sufficient responses from Bangladesh using MTurk, so we collected more than 100 responses by circulating the DPD questionnaire with our network in Bangladesh, i.e.,~manual sharing.  This is a diverse data collection compared to most of the previous works in the digital divide which collect responses from one country~\cite{redmiles2017digital,bonnefon2016social}. Previous works have shown that responses collected using MTurk in security and privacy studies provide a good representation of the general population~\cite{redmiles2019well}.

%
%
%
%
%

\section*{\textbf{DPD and Social Dilemma}}\label{sec:sec_3}

This section discusses the results of the DPD questionnaire. First, the DPD problem will be presented by analyzing the answers of respondents living in the United States, Germany, Bangladesh, and India. Second, the social dilemma of DPD will be presented.

\subsection*{\textbf{DPD at countries}}

\begin{figure*}
	\begin{centering}
		\includegraphics[width=1\linewidth,trim=0.5cm 0.5cm 1cm 2cm]{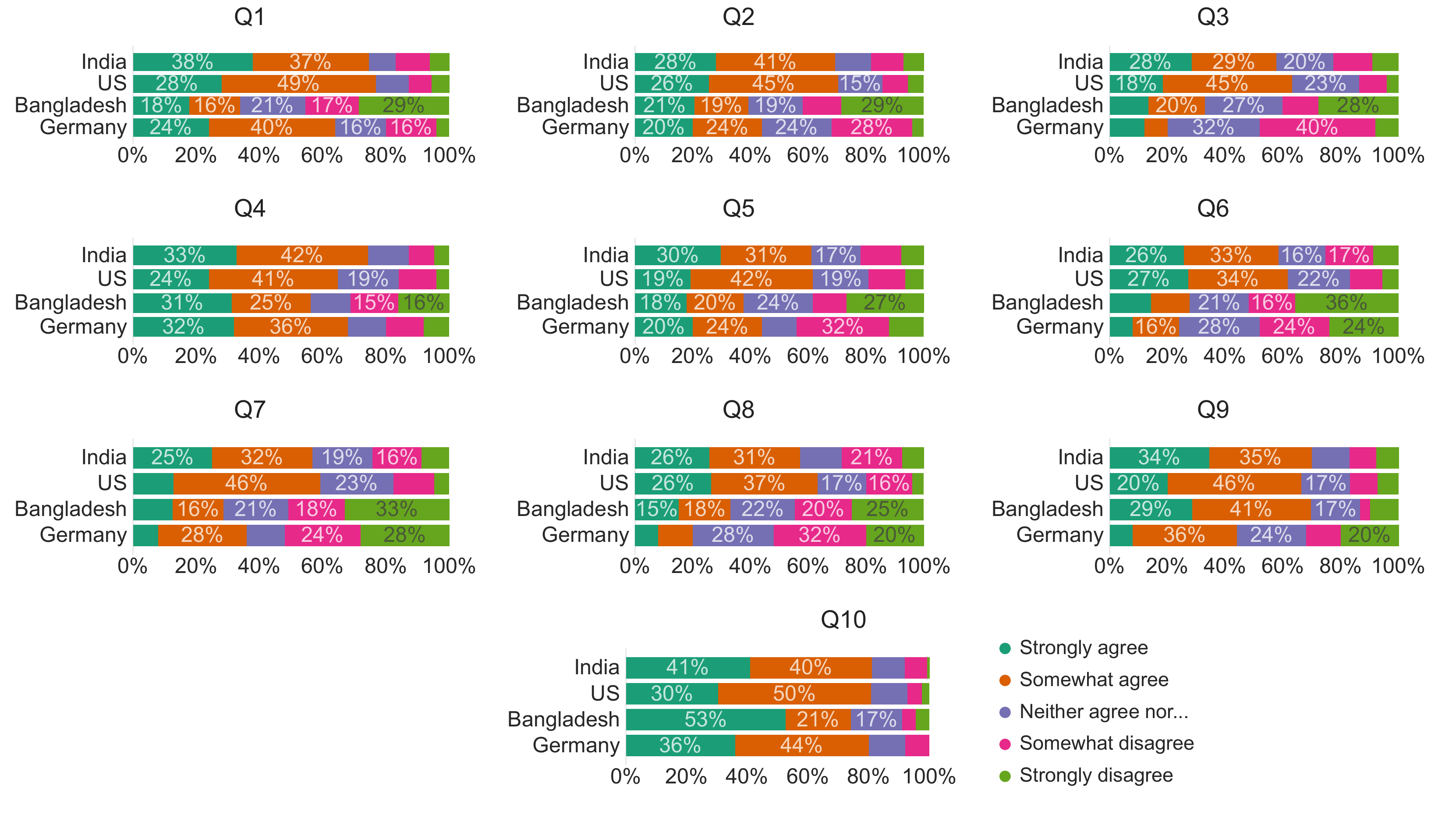}
		\par\end{centering}
	
	\caption{Percentages of the responses to Q1-Q10 divided based on the respondent's current country of living.\label{fig:country_privacy}}
\end{figure*}

Figure~\ref{fig:country_privacy} shows the percentage of responses of Q1-Q10 for respondents living in the United States, Germany, Bangladesh, and India. We find that the privacy concerns of individuals cannot be justified using theories of popular cultural dimensions, e.g.,~the Hofstede's model~\cite{hofstede2011dimensionalizing}. Moreover, respondents residing in Germany and Bangladesh have expressed more concerns about their privacy compared to those residing in the United States and India (Q1-Q8). This finding suggests the notion of privacy does not depend just on the simplistic cultural orientation of individualism vs. collectivism, rather it may encompass other internal or external factors at the national level.

\subsection*{\textbf{Social dilemma of DPD}}

We find that most of the respondents (80\%, 81\%,  74\%, and 80\% of the respondents in the United States, India, Bangladesh, and Germany, respectively) think that governments should enforce restrict privacy policies on domestic companies and organizations that collect their personal data in all cases, regardless of the economic, employment, and crime prevention benefits of the data collection (Q10). On the contrary, 66\%, 69\%, 70\%, and 44\% of the respondents in the United States, India, Bangladesh, and Germany think that governments should allow domestic companies and organizations collecting personal data on individuals residing outside their countries if there are economical, employment, and crime prevention benefits (Q9).
\section*{\textbf{Closing the Digital Privacy Divide}}\label{sec:sec_4}

Inequality in privacy preservation reinforces social inequality. In this section, we provide recommendations for closing the DPD gap in three main dimensions (educational, legislative, and technical dimensions).

\subsection*{\textbf{Education and awareness initiatives on privacy rights}}

Many people may not be fully aware of the pervasive nature of existing sensing technologies in ICT. Previous research has shown that users may not use privacy tools when they lack a clear understanding of their benefits and architectures (mental models)~\cite{renaud2014doesn}. Educational interventions and awareness programs and campaigns  must be initiated to increase the awareness of privacy protection and privacy rights amongst the citizens. Online services provided by governments must include an explicit description of the privacy rights. 

Knowledge and awareness on disclosures, marketing materials, loopholes related to online privacy on social networking sites, malware, adware to collect personal information about the users should be widely disseminated to prevent possible breaches.

\subsection*{\textbf{Privacy legislation and policy}}

There are various legislation policies all over the world, e.g.,~the General Data Protection Regulation (European Union), Privacy Act 1988 (Australia), and  General Personal Data Protection Law (Brazil). The differences in privacy legislation laws result in various interpretations of the privacy rights and the acceptable levels of privacy protection. A multi-stakeholder view of privacy, capturing local contextual and cultural interpretation of privacy, should be reflected in the policy. Questions should be raised about to what extent the government should invade its citizens' privacy, showing the pretext of national security. Potential boundaries and trade-offs between individual rights and security must be drawn carefully by examining its short- and long-term implications to individuals and society. Associated laws should keep up with the fast-paced changes of technological development to prevent possible gaps and confusion.

\subsection*{\textbf{Accessible privacy tools}}

A critical aspect of solving the DPD gap is in providing accessible and easy-to-use privacy tools for both users and service providers. Service providers need ICT tools to meet the privacy rights in IoT, networking, machine learning, and cloud computing. Users require usable tools for verifying the compliance of the service providers with the privacy regulation. Both sets of tools will contribute to closing the DPD gap.
\section*{\textbf{Future Directions}}\label{sec:future_work}

This section suggests interesting research directions on DPD that can be pursued in the future.

\subsection*{\textbf{DPD, socio-demographic, and socio-economic status}}

People from different socio-economic backgrounds can experience different security and privacy problems and have different  expectations about their privacy-related issues~\cite{redmiles2017digital}. For example, some people are more willing to share their personal data to make more friends while others are wary of sharing their personal data. There is an apprehension that generation Y, growing in the digital age may view the privacy issue differently than their previous generation or older counterpart. Such perception of privacy is also likely to be the  result of locally situated work practices and socially negotiated realities that impact privacy behavior in those settings. Future research can use the theoretical approach of ``situating culture'' for a better understanding of contextualism and cross-cultural issues in privacy perception among people of different countries and regions.


\subsection*{\textbf{COVID-19 and DPD}}

An interesting research direction is studying the impact of the COVID-19 pandemic in DPD. Previous studies have discussed the impacts of COVID-19 in telehealth access~\cite{ramsetty2020impact} and remote educational technologies~\cite{lai2021revisiting}. In disease outbreaks, contact tracing systems are used to monitor interactions with infected patients~\cite{ahmed2020survey}. Even though tracing applications are important for monitoring infection trends,  tracing technologies have major privacy risks~\cite{tedeschi2021iotrace}. An interesting future research direction is analyzing the impact of disease outbreaks on the acceptance of higher privacy risks and the trade-offs between privacy compromise and potential benefits. 

\subsection*{\textbf{Mixed method study of DPD}}

A mixed-method  study followed by a qualitative inquiry should be conducted to provide a deeper understanding and rich insight of complex socio-cultural and socio-political dimensions influencing the perception of privacy and to complement quantitative survey data. Such an approach helps to explain the underlying causes of change in citizens' behavior over time, the role and impact of power relations between governments and citizens, and trust in ICT services.

\section*{\textbf{Conclusion}}\label{sec:conclusion}

In this article, we have discussed the problem of the digital privacy divide~(DPD). We have found that DPD does not depend on Hofstede's cultural orientation of the countries. We have also found that most respondents are concerned about their digital privacy while supporting the data collection about individuals residing outside their countries if there are economic, employment, and crime prevention benefits.

Breaches of privacy pose a great threat to individual freedom and dignity in the modern world, which has far-reaching ramifications to create a just society. While privacy is being a highly personalized issue and differs from person to person, a common trend and characteristic based on local circumstances can be generalized. Future studies must be undertaken to unearth those unexplained factors in order to define the DPD issues more objectively and narrow the DPD gaps between countries.

Due to the increasingly global nature of ICT access and use, a global consensus must be reached to preserve human dignity and privacy rights for better societies and to narrow the digital privacy gaps.

\balance
\bibliographystyle{IEEEtran}
\bibliography{reference}

\begin{thebibliography}{10}
\providecommand{\url}[1]{#1}
\csname url@samestyle\endcsname
\providecommand{\newblock}{\relax}
\providecommand{\bibinfo}[2]{#2}
\providecommand{\BIBentrySTDinterwordspacing}{\spaceskip=0pt\relax}
\providecommand{\BIBentryALTinterwordstretchfactor}{4}
\providecommand{\BIBentryALTinterwordspacing}{\spaceskip=\fontdimen2\font plus
\BIBentryALTinterwordstretchfactor\fontdimen3\font minus
  \fontdimen4\font\relax}
\providecommand{\BIBforeignlanguage}[2]{{%
\expandafter\ifx\csname l@#1\endcsname\relax
\typeout{** WARNING: IEEEtran.bst: No hyphenation pattern has been}%
\typeout{** loaded for the language `#1'. Using the pattern for}%
\typeout{** the default language instead.}%
\else
\language=\csname l@#1\endcsname
\fi
#2}}
\providecommand{\BIBdecl}{\relax}
\BIBdecl

\bibitem{ramsetty2020impact}
A.~Ramsetty and C.~Adams, ``Impact of the digital divide in the age of
  {COVID-19},'' \emph{Journal of the American Medical Informatics Association},
  vol.~27, no.~7, pp. 1147--1148, April 2020.

\bibitem{lai2021revisiting}
J.~Lai and N.~O. Widmar, ``Revisiting the digital divide in the {COVID-19}
  era,'' \emph{Applied Economic Perspectives and Policy}, vol.~43, no.~1, pp.
  458--464, October 2021.

\bibitem{scheerder2017determinants}
A.~Scheerder, A.~van Deursen, and J.~van Dijk, ``Determinants of {I}nternet
  skills, uses and outcomes. {A} systematic review of the second-and
  third-level digital divide,'' \emph{Telematics and Informatics}, vol.~34,
  no.~8, pp. 1607--1624, December 2017.

\bibitem{gdpr2016general}
{European Parliament} and {Council of the European Union}, ``General data
  protection regulation ({GDPR}),'' \url{https://gdpr-info.eu/}, 2016, online;
  accessed 2 August 2021.

\bibitem{hofstede2011dimensionalizing}
G.~Hofstede, ``Dimensionalizing cultures: {T}he {H}ofstede model in context,''
  \emph{Online Readings in Psychology and Culture}, vol.~2, no.~1, pp.
  2307--0919, January 2011.

\bibitem{kement2021holistic}
C.~E. Kement, B.~Tavli, H.~Gultekin, and H.~Yanikomeroglu, ``Holistic privacy
  for electricity, water, and natural gas metering in next generation smart
  homes,'' \emph{IEEE Communications Magazine}, vol.~59, no.~3, pp. 24--29,
  March 2021.

\bibitem{liu2021machine}
B.~Liu, M.~Ding, S.~Shaham, W.~Rahayu, F.~Farokhi, and Z.~Lin, ``When machine
  learning meets privacy: {A} survey and outlook,'' \emph{ACM Computing
  Surveys}, vol.~54, no.~2, pp. 1--36, April 2021.

\bibitem{redmiles2017digital}
E.~M. Redmiles, S.~Kross, and M.~L. Mazurek, ``Where is the digital divide? {A}
  survey of security, privacy, and socioeconomics,'' in \emph{Proceedings of
  the International Conference on Human Factors in Computing Systems}, 2017,
  pp. 931--936.

\bibitem{saris2014design}
W.~E. Saris and I.~N. Gallhofer, \emph{Design, evaluation, and analysis of
  questionnaires for survey research}.\hskip 1em plus 0.5em minus 0.4em\relax
  John Wiley \& Sons, 2014.

\bibitem{bonnefon2016social}
J.-F. Bonnefon, A.~Shariff, and I.~Rahwan, ``The social dilemma of autonomous
  vehicles,'' \emph{Science}, vol. 352, no. 6293, pp. 1573--1576, June 2016.

\bibitem{zhang2014speeding}
C.~Zhang and F.~Conrad, ``Speeding in web surveys: {T}he tendency to answer
  very fast and its association with straightlining,'' \emph{Survey Research
  Methods}, vol.~8, no.~2, pp. 127--135, July 2014.

\bibitem{redmiles2019well}
E.~M. Redmiles, S.~Kross, and M.~L. Mazurek, ``How well do my results
  generalize? {C}omparing security and privacy survey results from {MTurk},
  web, and telephone samples,'' in \emph{Proceedings of the International
  Symposium on Security and Privacy}.\hskip 1em plus 0.5em minus 0.4em\relax
  IEEE, 2019, pp. 1326--1343.

\bibitem{renaud2014doesn}
K.~Renaud, M.~Volkamer, and A.~Renkema-Padmos, ``Why doesn’t {J}ane protect
  her privacy?'' in \emph{Proceedings of the International Symposium on Privacy
  Enhancing Technologies Symposium}.\hskip 1em plus 0.5em minus 0.4em\relax
  Springer, 2014, pp. 244--262.

\bibitem{ahmed2020survey}
N.~Ahmed, R.~A. Michelin, W.~Xue, S.~Ruj, R.~Malaney, S.~S. Kanhere,
  A.~Seneviratne, W.~Hu, H.~Janicke, and S.~K. Jha, ``A survey of {COVID}-19
  contact tracing apps,'' \emph{IEEE Access}, vol.~8, pp. 134\,577--134\,601,
  July 2020.

\bibitem{tedeschi2021iotrace}
P.~Tedeschi, S.~Bakiras, and R.~Di~Pietro, ``{IoTRace}: {A} flexible,
  efficient, and privacy-preserving {IoT}-enabled architecture for contact
  tracing,'' \emph{IEEE Communications Magazine}, vol.~59, no.~6, pp. 82--88,
  June 2021.

\end{thebibliography}
\section*{\textbf{Biographies}}

\begin{IEEEbiographynophoto}
{Hamoud Alhazmi}
(halhazmi@ieee.org) [S'21] is working as a Research Assistant at the University of Canberra, ACT, Australia. He graduated with a Master's degree in Cybersecurity and B.Eng. degree in Network \& Software Engineering with (first-class honors) from the University of Canberra in 2021 and 2020, respectively. His current research interests are in computer vision, machine learning, and cybersecurity. He worked as an Assistant IT Manager during his studies in Canberra.
\end{IEEEbiographynophoto}

\begin{IEEEbiographynophoto}
{Ahmed Imran}
(ahmed.imran@canberra.edu.au) is an Information Systems researcher at the University of Canberra with special interests in the strategic use of IT, eGovernment, and socio-cultural impacts of ICT. His vast experience as an IT manager as well as his work in developing countries became invaluable for research, and in understanding and providing a rich insight into the socio-cultural context through multiple lenses, resulting in interdisciplinary research opportunities. His research has proven to bring real-world applications to the table, something that cemented its importance and relevance in the eyes of the research community. This recognition was further reflected through the award of the prestigious Australian National University Vice Chancellor's award in 2010 followed by numerous invitations to international and national forums/universities.
\end{IEEEbiographynophoto}

\begin{IEEEbiographynophoto}
{Mohammad Abu Alsheikh}
(mabualsh@ieee.org) [S'14--M'17] is an Associate Professor and ARC DECRA Fellow at the University of Canberra, Australia. He was a Postdoctoral Researcher at the Massachusetts Institute of Technology, USA. He designs and creates novel privacy-preserving Internet of things systems that leverage both machine learning and convex optimization with applications in people-centric sensing, human activity recognition, and smart cities. His doctoral research at Nanyang Technological University, Singapore was focused on optimizing the data collection in wireless sensor networks. After graduating with a B.Eng. degree in computer systems from Birzeit University, Palestine, he worked as a software engineer at a digital advertising start-up and Cisco.
\end{IEEEbiographynophoto}

\vfill

\end{document}